\documentclass[aps,prd,superscriptaddress,11pt]{revtex4-2}
\usepackage{amsmath,amssymb,amsfonts}
\usepackage{graphicx}
\usepackage[english]{babel}
\usepackage{physics}
\usepackage[dvipsnames]{xcolor}
\usepackage{aas_macros}
\usepackage{soul}
\usepackage{orcidlink}
\usepackage{multirow}
\usepackage{array}
\usepackage[normalem]{ulem}

\usepackage{lineno}
\usepackage{comment}

\newcommand{\comoving}[1]{{\mathbf #1}}

\usepackage{hyperref}
\hypersetup{
    colorlinks=true,
    linkcolor=blue,
    citecolor=blue,
    filecolor=magenta,      
    urlcolor=blue,
    pdftitle={Cermenati+,2025},
    pdfpagemode=FullScreen,
    }
    
\graphicspath{{./figs/}}

\begin{document}
\title{Scrutinizing the cosmogenic origin of the KM3-230213A event: \\ A Multimessenger Perspective}

\author{Alessandro Cermenati~\orcidlink{0009-0002-7015-2977}}
\email[]{alessandro.cermenati@gssi.it}
\affiliation{Gran Sasso Science Institute (GSSI), Viale Francesco Crispi 7, 67100 L’Aquila, Italy}
\affiliation{INFN-Laboratori Nazionali del Gran Sasso (LNGS), via G. Acitelli 22, 67100 Assergi (AQ), Italy}

\author{Antonio Ambrosone~\orcidlink{0000-0002-9942-1029}}
\email[]{antonio.ambrosone@gssi.it}
\affiliation{Gran Sasso Science Institute (GSSI), Viale Francesco Crispi 7, 67100 L’Aquila, Italy}
\affiliation{INFN-Laboratori Nazionali del Gran Sasso (LNGS), via G. Acitelli 22, 67100 Assergi (AQ), Italy}

\author{Roberto Aloisio~\orcidlink{0000-0003-0161-5923}}
\affiliation{Gran Sasso Science Institute (GSSI), Viale Francesco Crispi 7, 67100 L’Aquila, Italy}
\affiliation{INFN-Laboratori Nazionali del Gran Sasso (LNGS), via G. Acitelli 22, 67100 Assergi (AQ), Italy}

\author{Denise Boncioli~\orcidlink{0000-0003-1186-9353}}
\affiliation{Università degli Studi dell’Aquila, Dipartimento di Scienze Fisiche e Chimiche, Via Vetoio, 67100, L’Aquila, Italy}
\affiliation{INFN-Laboratori Nazionali del Gran Sasso (LNGS), via G. Acitelli 22, 67100 Assergi (AQ), Italy}

\author{Carmelo Evoli~\orcidlink{0000-0002-6023-5253}}
\affiliation{Gran Sasso Science Institute (GSSI), Viale Francesco Crispi 7, 67100 L’Aquila, Italy}
\affiliation{INFN-Laboratori Nazionali del Gran Sasso (LNGS), via G. Acitelli 22, 67100 Assergi (AQ), Italy}

\newcommand{\CE}[1]{{\color{red!70!black}#1}}

\date{\today}

\begin{abstract}
The recent detection of the neutrino event KM3-230213A ($\sim$~220 PeV) by the KM3NeT/ARCA telescope—the most energetic ever observed—could represent the long-awaited evidence for a cosmogenic origin, arising from the interaction of an ultra-high-energy cosmic ray with background photons. Its secure confirmation would mark a major advance in high-energy astrophysics.
We perform a self-consistent multimessenger transport calculation of protons and their secondary $\gamma$-rays and neutrinos from cosmologically evolving sources, confronting predictions with data from the Pierre Auger Observatory, IceCube, KM3NeT and the Fermi-LAT isotropic $\gamma$-ray background. A steep sub-ankle proton component saturates the diffuse $\gamma$-ray background and is disfavoured, whereas a hard proton spectrum extending beyond $10^{20}$~eV with evolution $\propto (1+z)^3$ reproduces KM3-230213A without violating any limits. This scenario requires a proton fraction $\lesssim 10$\% at $3\times 10^{19}$~eV and excludes faster-evolving sources. 
Joint UHE-neutrino and $\gamma$-ray observations thus sharpen constraints on extragalactic cosmic-ray sources and set targets for AugerPrime and next-generation neutrino telescopes.
\end{abstract}

\maketitle

\newpage

%%%%%%%%%% BEGIN SECTION %%%%%%%%%%
\section*{Main text}

The recent detection by the Cubic Kilometer Neutrino Telescope (KM3NeT) of KM3-230213A -- the most energetic astrophysical neutrino ever observed -- marks a turning point in astroparticle physics~\cite{KM3NeT:2025npi}. 
This extraordinary track-like event, produced by a neutrino traversing the Earth and generating a Cherenkov-light-emitting muon, triggered some 3,000 photomultipliers. 
Its reconstructed energy, $220_{-148}^{+2380}~\rm PeV$ (90\% CL), ventures deep into the ultra-high-energy regime, corresponding to parent particle energies typical of the so-called ``ankle'' in the cosmic-ray spectrum.

Neutrinos such as this are thought to originate in the decay chain of pions produced when cosmic rays interact with photons within their source environment (\emph{astrophysical neutrinos}), as well as with photons encountered during propagation through extragalactic space (\emph{cosmogenic neutrinos}), such as the cosmic microwave background (CMB) and the extragalactic background light (EBL)~\cite{Serpico:2023yag,Boncioli:2023gbl}. 
These same processes also yield extremely energetic gamma rays, which are further reprocessed, via inverse Compton scattering and pair production, to GeV energies by electromagnetic cascades~\cite{Berezinsky:2016feh}.
This realization has then sparked an ambitious research effort that aims to exploit simultaneous observations of high-energy neutrinos and the diffuse gamma-ray background, furnishing complementary—and potentially decisive—insights into the hadronic processes driving ultra-high-energy cosmic-ray (UHECR) production and propagation.

For the first time, the observation of KM3-230213A allows us to probe the properties of cosmic rays at the highest energies through a messenger immune to magnetic deflection and absorption. 
The neutrino’s energy corresponds to parent protons firmly in the UHECR regime, directly linking the event to the long-standing puzzle of cosmic-ray origins above the \emph{ankle}, a striking hardening of the all-particle cosmic-ray spectrum at roughly \( 4 \times 10^{18} \)~eV~\cite{PierreAuger:2020qqz}.

Yet, the cosmogenic interpretation of this event remains unsettled, hinging critically on the mass composition of UHECRs at the highest energies: recent measurements reveal a trend toward heavier nuclei with a narrow dispersion among mass groups, although a minor proton fraction still falls within current upper limits~\cite{PierreAuger:2023kjt,PierreAuger:2023xfc}.

This issue is, however, central. On one hand, the amount of protons in the UHECR flux is crucially connected to the resulting neutrino and photon spectra, because the energy threshold for photopion production scales linearly with nuclear mass. Hence, protons govern the energy range and shape of the neutrino and photon fluxes~\cite{Aloisio:2015ega,Heinze:2015hhp,AlvesBatista:2018zui,Heinze:2019jou,Ehlert:2023btz}. On the other hand, the interactions limit the visible Universe for UHECRs, which consequently cannot be observed from distance sources. Neutrinos and photons, therefore, provide much greater discriminating power regarding the cosmological evolution of UHECR sources~\cite{vanVliet:2019nse,PierreAuger:2019ens,PierreAuger:2023mid,Muzio:2023skc,IceCube:2025ezc}.

In this work, we critically examine the cosmogenic origin hypothesis for KM3-230213A~\cite{KM3NeT:2025vut,Li:2025tqf,Zhang:2025abk,Muzio:2025gbr}. 
We adopt a multimessenger strategy, jointly analysing UHECR, neutrino, and gamma-ray data. This includes the proton spectrum inferred from the Pierre Auger Observatory~\cite{PierreAuger:2023kjt, PierreAuger:2023xfc}, neutrino fluxes and limits from Auger~\cite{PierreAuger:2023pjg}, IceCube~\cite{IceCube:2025ezc}, and KM3NeT~\cite{KM3NeT:2025ccp}, and the Fermi diffuse gamma-ray background~\cite{Fermi-LAT:2014ryh}. 
%
%\OLD{Upper limits on the ultra-high-energy photon flux~\cite{PierreAuger:2024ayl} could also be considered; however, the flux of ultra-high-energy photons critically depends on the local UHECR density, and it is almost independent of the cosmological evolution of UHECR sources.} 
%
By solving cosmological transport equations for protons, neutrinos, and photons, and considering two source populations below and above the ankle, we quantify the power of current neutrino and photon data to constrain the energy range and cosmological evolution of proton sources. Our results set new, stringent limits on the possible origins of the highest-energy neutrinos, and demonstrate the power of multimessenger exploration of the extreme Universe.
%%%%%%%%%% END SECTION %%%%%%%%%%

\begin{figure}[t!]
\centering
\includegraphics[width=0.49\textwidth]{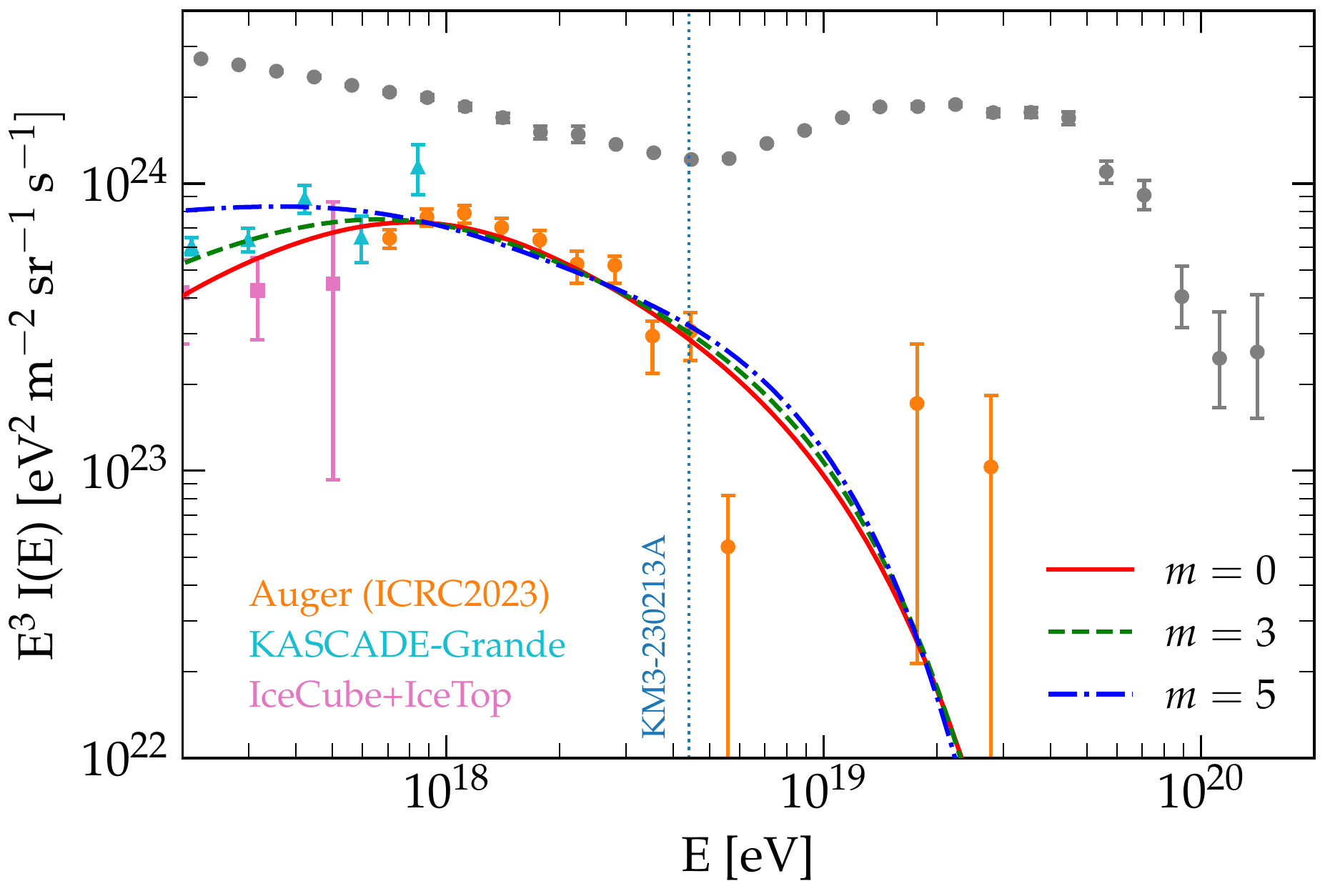}
\hspace{\stretch{1}}
\includegraphics[width=0.49\textwidth]{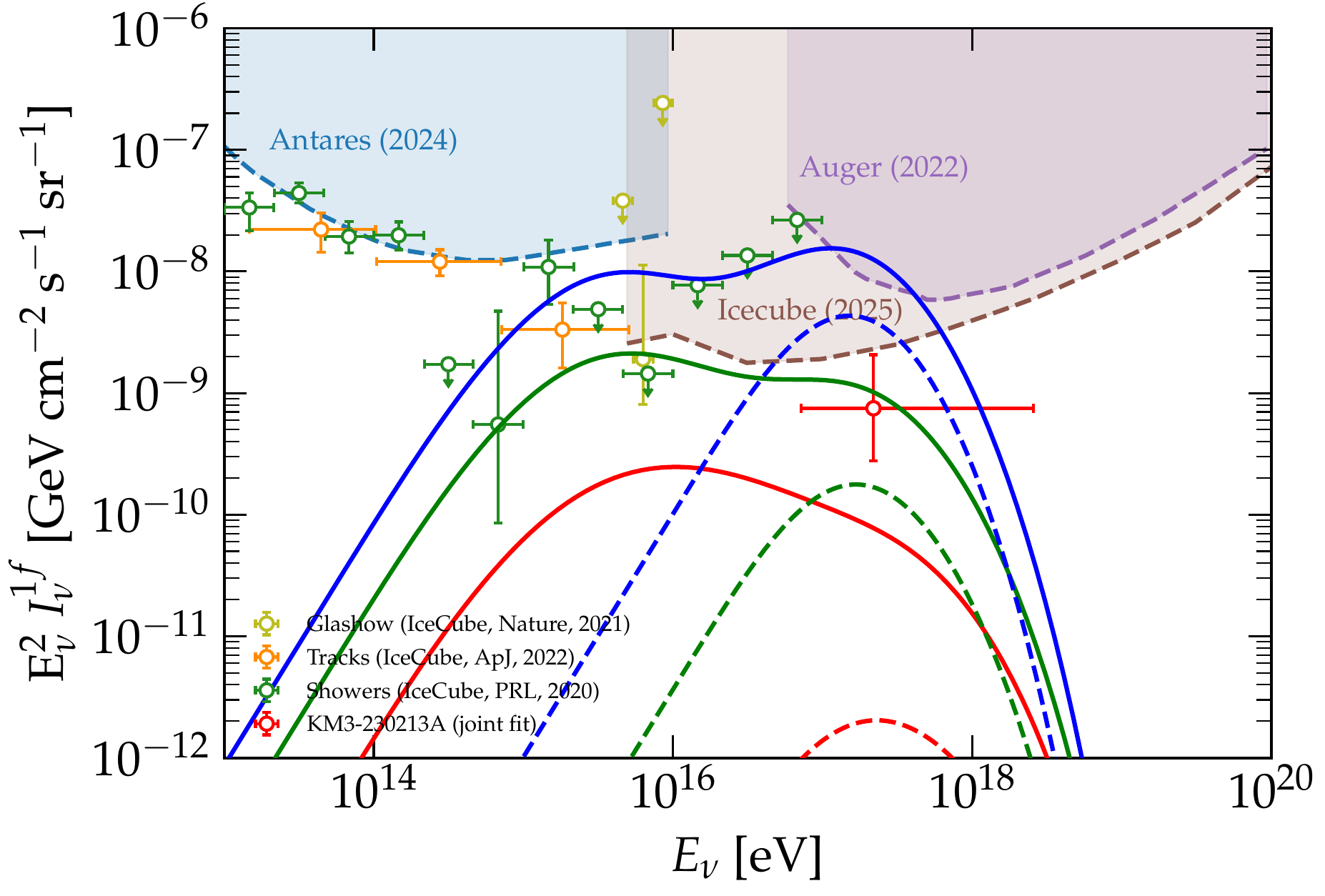}
\caption{
\textit{Left:} UHECR fluxes, multiplied by $E^3$. Colored lines represent the expected proton fluxes obtained using the parameter set in Tab.~\ref{tab:set_parameter}. The grey and orange (circles) points indicate, respectively, the Auger all-particle spectrum~\cite{PierreAuger:2021hun} and the Auger proton spectrum as inferred with Sibyll2.3d (using the proton fraction from~\cite{PierreAuger:2023xfc}). For comparison, the cyan (triangles) and purple (squares) points display the proton spectra measured by KASCADE-Grande~\cite{KASCADEGrande:2017gtn} and IceCube/IceTop~\cite{IceCube:2019hmk}, both inferred using the same hadronic model. The vertical blue (dotted) line marks the proton energy corresponding to the KM3NeT neutrino event.
\textit{Right:} Single-flavour neutrino fluxes, multiplied by $E^2$, corresponding to the proton fluxes shown in the left panel and using the same color scheme. Solid lines indicate the total contribution from CMB and EBL, while dashed lines show the CMB-only contribution. Data points correspond to various IceCube measurements~\cite{IceCube:2020acn,Abbasi:2021qfz,IceCube:2021rpz}; the red point marks the joint flux measured by KM3NeT, Auger, and IceCube associated with the KM3-230213A event~\cite{KM3NeT:2025ccp}.
Shaded regions correspond to upper limits from ANTARES (95\% CL~\cite{ANTARES:2024ihw}), Auger (90\% CL~\cite{PierreAuger:2023pjg}) and IC-EHE (90\% CL~\cite{IceCube:2025ezc}).}
\label{LE}
\end{figure}

%%%%%%%%%% BEGIN SECTION %%%%%%%%%%
\section{Results and Discussion}
\label{sec::Results}

We model the UHECR flux as arising from two distinct source populations. The first, which we designate as the low-energy (LE) population, accounts for the measured proton flux below the ankle. The second, referred to as the high-energy (HE) population, contributes above the ankle with a harder spectrum. For the LE population, we normalize the source emissivity to match the proton spectrum measured by Auger adopting Sibyll2.3d as the interaction model for simulating atmospheric showers~\citep{PierreAuger:2023kjt,PierreAuger:2023xfc}. For the HE population, the normalization is set to be consistent with a 10\% upper limit of the all-particle spectrum at approximately $3 \times 10^{19}~\rm eV$, as suggested by mass fraction analyses~\cite{PierreAuger:2023xfc,PierreAuger:2024flk}.

To fully explore the parameter space, we consider various choices for the spectral shape, maximum energy of the proton fluxes, and cosmological source evolution, the latter parametrized as a power law $(1+z)^m$ up to a maximum redshift of $z_{\rm max} = 5$. Our approach, which separates proton sources into LE and HE populations, is similar to that of~\cite{Ehlert:2023btz}. %, although the proton fraction has not been explicitly used in that work.

\begin{figure}[t!]
\centering
\includegraphics[width=0.49\textwidth]{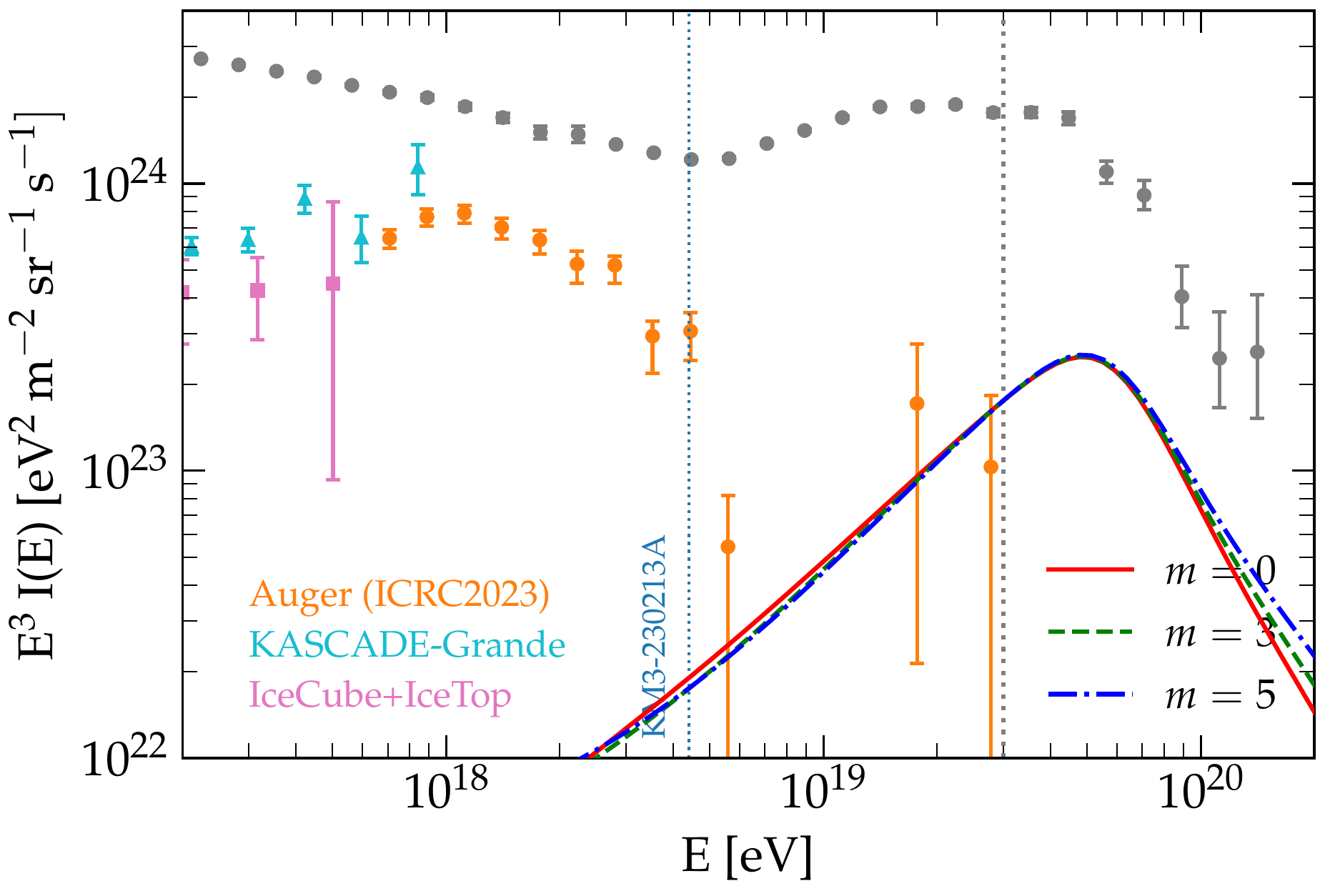}
\hspace{\stretch{1}}
\includegraphics[width=0.49\textwidth]{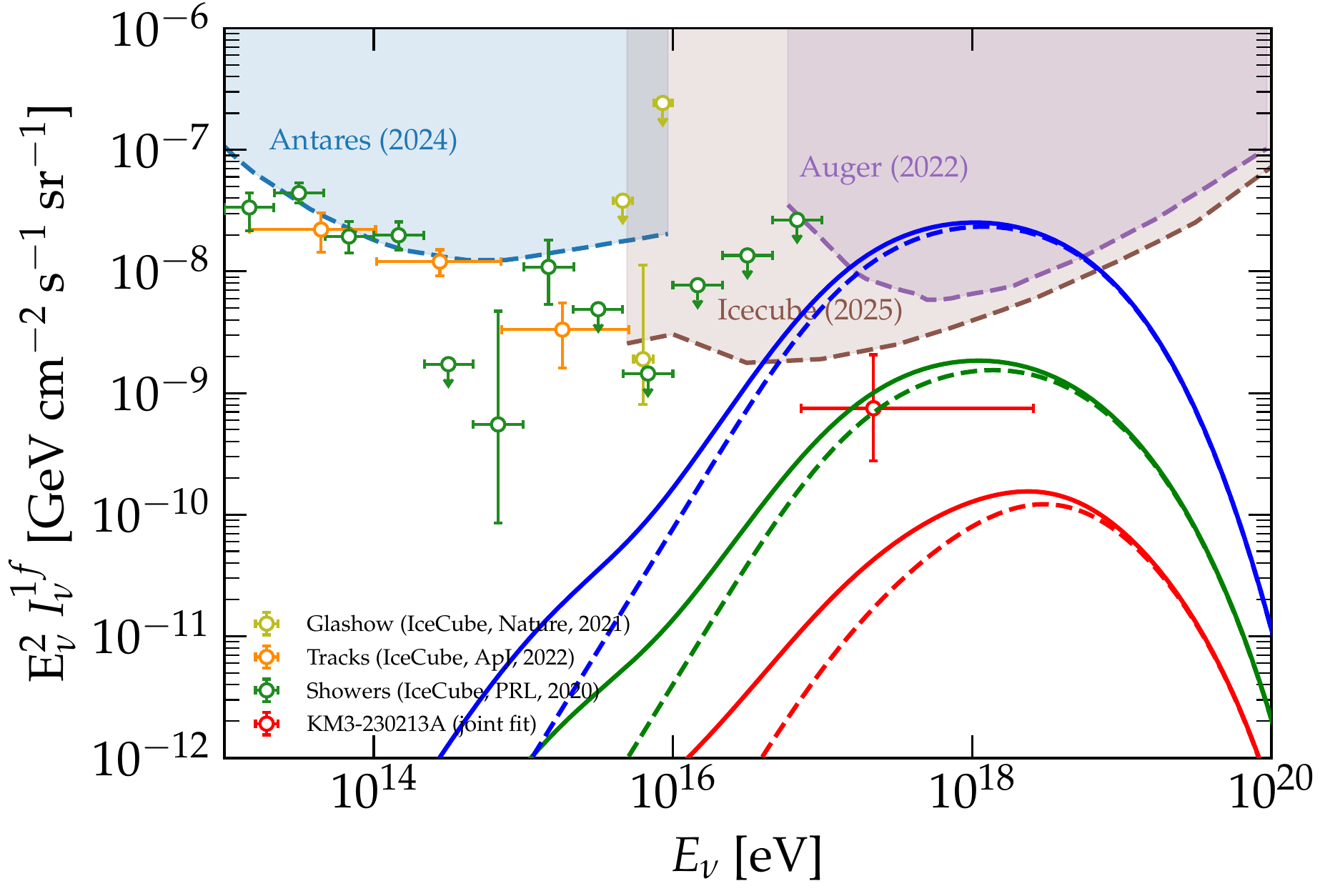}
\caption{Same as Fig.~\ref{LE}, but for the high-energy (HE) population, with model parameters given in Tab.~\ref{tab:set_parameter}. 
Left: UHECR proton fluxes (multiplied by $E^3$) predicted for the HE population, shown together with the same experimental datasets as in Fig.~\ref{LE}. The grey line marks the reference energy where the model proton flux is normalized to 10\% of the all-particle flux.
Right: Corresponding single-flavour neutrino fluxes (multiplied by $E^2$) for the HE population, using the same color scheme and conventions as in Fig.~\ref{LE}.}
\label{HE}
\end{figure}

Table~\ref{tab:set_parameter} summarizes the parameter choices adopted for the proton source models in both the LE and HE ranges. For each parameter set, Figures~\ref{LE} and~\ref{HE} display the resulting proton spectra at $z = 0$ (left panels) alongside the corresponding cosmogenic neutrino spectra (right panels).

We compare our predictions about cosmogenic neutrinos with the IceCube neutrino diffuse fluxes~\cite{IceCube:2020acn,Abbasi:2021qfz,IceCube:2021rpz}, and with the joint KM3NeT, IceCube and Auger neutrino flux estimate from the KM3-230213A event, as reported in~\cite{KM3NeT:2025ccp}. Dashed lines in the neutrino spectra show the contribution from interactions of protons with the CMB only, while solid lines show the total (CMB+EBL).

For the LE population, the neutrino spectra show a two-bump feature corresponding to protons interacting with, respectively, visible (infrared) light at lower (higher) energies. 
In fact, the maximum energy of protons lies below the threshold for photopion production on the CMB, making the corresponding neutrino contribution subdominant with respect to the EBL; however, this contribution increases with stronger source evolution, as the CMB temperature and density rise at higher redshifts. We find that a moderate source evolution ($m \sim 3$) can account for the neutrino flux associated with the KM3-230213A event. Stronger evolution, however, would lead to neutrino fluxes that are already in tension with IceCube measurements at PeV energies and exceed the limits set by Auger at energies above $10^{17}$~eV.

For the HE population (see Figure~\ref{HE}), cosmogenic neutrinos are predominantly produced through interactions with the CMB due to both the higher maximum energies and harder spectral indices. As with the LE case, an evolution parameter of $m \sim 3$ provides good agreement with the flux reported for KM3-230213A. In contrast, stronger evolution would again result in neutrino fluxes that surpass the constraints from Auger and IceCube—though in this scenario, the predicted emission extends to even higher energies and does not contribute significantly to the PeV flux.

\begin{figure}[t!]
\centering
\includegraphics[width=0.49\textwidth]{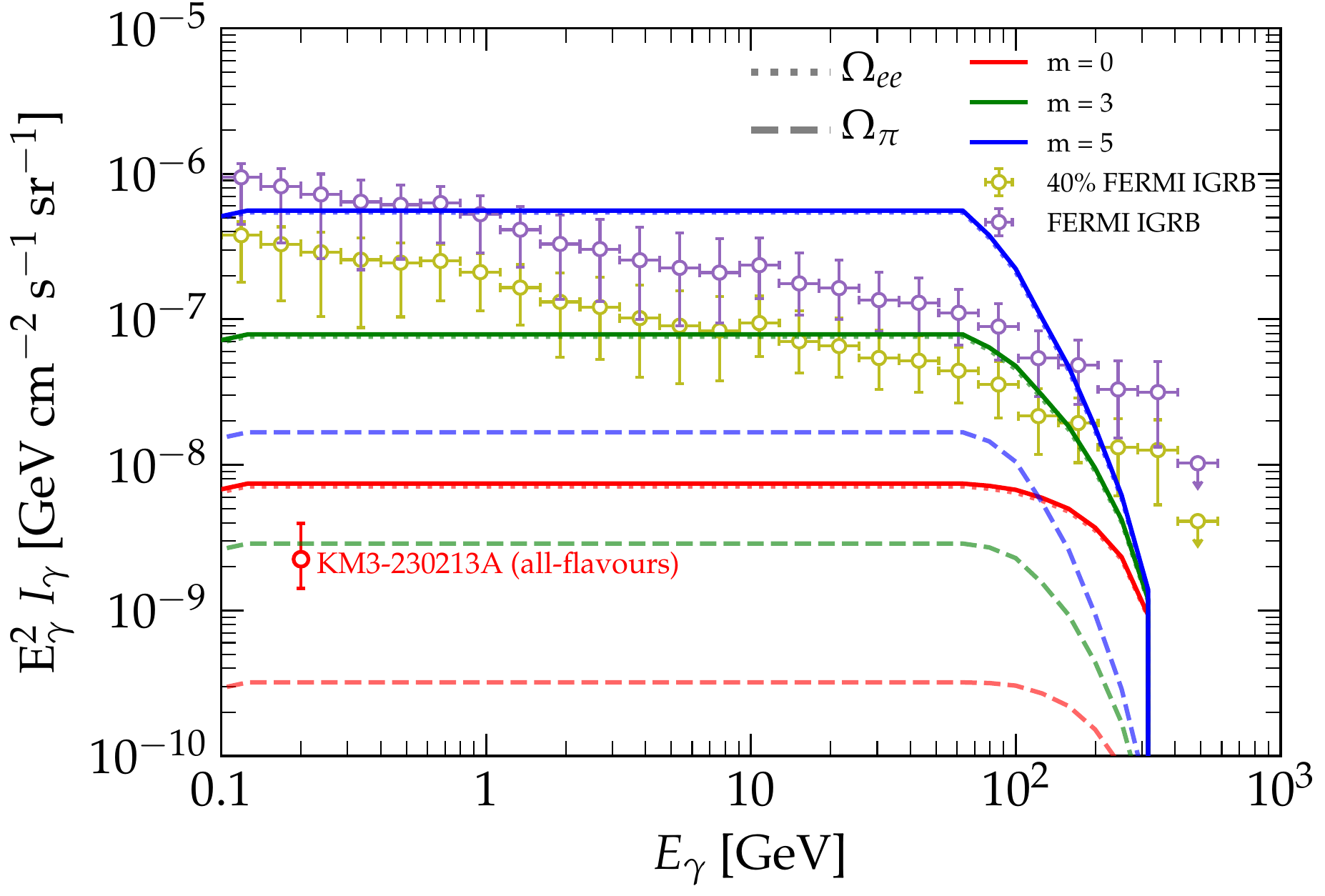}
\hspace{\stretch{1}}
\includegraphics[width=0.49\textwidth]{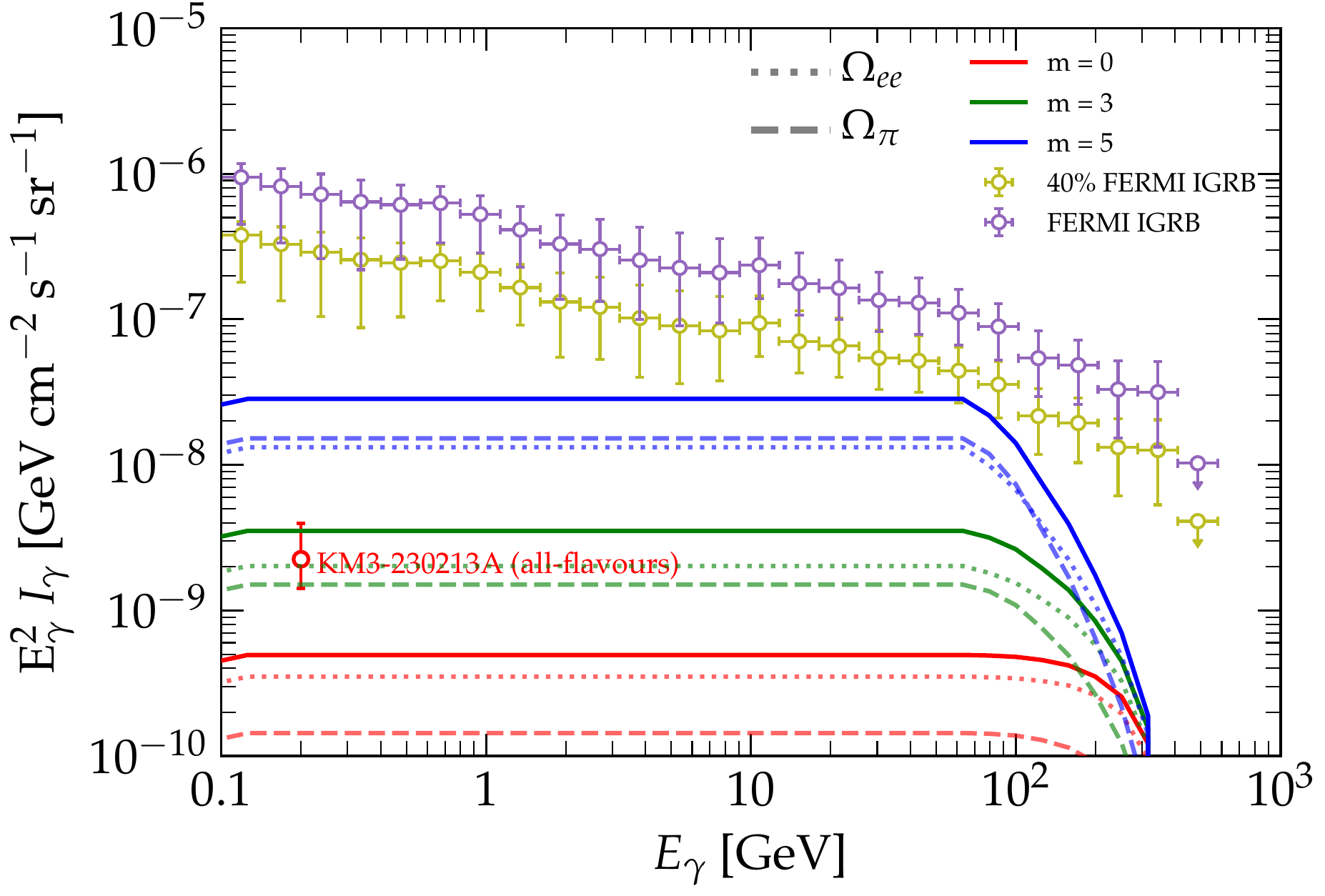}
\caption{Gamma-ray fluxes multiplied by $E^2$, shown using the same color scheme as in Fig.~\ref{LE}. The left panel corresponds to the low-energy (LE) population, and the right panel to the high-energy (HE) population. Solid lines represent the total gamma-ray flux (from proton pair production plus photopion production), while dotted and dashed lines indicate the individual contributions from pair production and photopion production, respectively.
Purple data points indicate the isotropic gamma-ray background (IGRB) measured by Fermi-LAT~\cite{Fermi-LAT:2014ryh}. Yellow data points show the same Fermi-LAT measurements, rescaled by a factor of 0.4 to account for the expected contributions from unresolved source populations~\cite{Ajello:2015mfa}.  For visual comparison, the red point marks the neutrino flux level (multiplied by $E^2$) associated with the joint KM3-230213A event.}
\label{gamma}
\end{figure}

While neutrino data alone would leave the cosmogenic origin degenerate between the LE and HE populations, with significant implications for the astrophysical origin of extragalactic protons, we now turn to their gamma-ray counterparts to break this degeneracy. 
An electromagnetic cascade is the chain reaction that ensues when a single high-energy $\gamma$-ray (or electron) propagates through radiation fields, matter, or magnetic fields, repeatedly converting into $e^+e^-$ pairs and lower-energy photons. In astrophysical environments, it efficiently channels the energy first absorbed from ultra-high-energy protons back into GeV-TeV $\gamma-$rays, reshaping the observable spectrum.
Our predictions for the diffuse gamma-ray fluxes from electromagnetic cascades are shown in Figure~\ref{gamma}, with the left and right panels corresponding to the LE and HE populations, respectively.

Focusing first on the LE population (left panel), we show the gamma-ray fluxes for the different source evolution parameters $m$ considered previously, and compare them with the total isotropic gamma-ray background (IGRB) measured by Fermi-LAT~\cite{Fermi-LAT:2014ryh}. The contributions from pair production and photopion production are shown separately (see Methods for details). 
The dashed curves represent the $\gamma$-ray flux generated solely by high-energy photons and pairs from photopion interactions. For a source spectral index $m = 3$, the resulting energy flux closely matches that of the neutrinos (red point), differing only by the neutral-to-charged pion ratio intrinsic to the photopion process. This correspondence reflects energy conservation in the resulting electromagnetic cascade, which redistributes the absorbed photon energy to lower frequencies and naturally leads to the universal $E^{-2}$ spectrum.

However, with the adopted soft proton spectra ($\gamma \gtrsim 2$) and maximum energies, an additional and substantial gamma-ray contribution arises from pair production on the CMB, with a threshold energy around $6 \times 10^{17}$~eV. The resulting energy density from this process exceeds that of photopion production by roughly an order of magnitude. As a result, the total predicted gamma-ray flux approaches or even saturates the IGRB measurements for source evolution parameters $m \gtrsim 3$.

This tension becomes even more pronounced when considering that a significant fraction (at least 60\%) of the IGRB is already attributed to unresolved sources from various populations~\cite{AlvesBatista:2018zui,Ambrosone:2020evo,Ajello:2015mfa,Lisanti:2016jub,Fermi-LAT:2015otn}. To illustrate this, we also plot the IGRB measured by Fermi-LAT scaled by a factor of 0.4, representing the maximum allowed contribution from truly diffuse processes. 

In contrast, the gamma-ray flux associated with the HE population is shown in the right panel of Fig.~\ref{gamma}. In this case, even for the highest values of source evolution, the predicted gamma-ray flux does not reach the observed IGRB. This is primarily because the contribution from pair production is strongly suppressed -- with respect to the LE case -- due to the hard spectral index of the HE population. The photopion-induced gamma-ray component, however, remains at a level comparable to that of the LE population, reflecting the fundamental requirement that the neutrino flux (produced exclusively through photopion production) must match the observed value.

Thus, gamma-ray observations impose a powerful constraint that strongly disfavors models with rapid source evolution in the LE population, as they would overproduce the diffuse gamma-ray background observed by Fermi-LAT. In contrast, these same observations remain fully compatible with the HE population, even for the highest source evolution scenarios, due to the suppressed pair production contribution. This complementary constraint from gamma rays therefore breaks the degeneracy left by neutrino observations alone, allowing us to distinguish between the two scenarios and favoring a cosmogenic neutrino origin linked to hard-spectrum, high-energy proton sources.

A few caveats of our analysis should be noted. The shape of the cutoff in the gamma-ray spectrum at $E_{\rm cutoff, \gamma} \sim 100$ GeV, where the reprocessed photons no longer follow the universal spectrum, has been implemented here using a standard approximation (see Methods and~\cite{Berezinsky:2016feh}). This approach assumes rapid development of electromagnetic cascades and a monochromatic EBL field at $\epsilon_{\rm EBL} \sim 0.7$~eV. In reality, a more accurate EBL distribution would produce a smoother cutoff, located at slightly lower energies. Nonetheless, our key conclusions, that strong source evolution ($m \gtrsim 3$) yields a diffuse gamma-ray flux in tension with observations, are derived primarily from the energy regime $E_\gamma \ll E_{\rm cutoff, \gamma}$, where the approximation is robust. Preliminary tests using a full cascade treatment and a realistic EBL confirm our main results~\footnote{A.~Cermenati et al., 2025, \textit{in preparation}}, supporting the validity of our simplified approach.

Another simplifying choice is our fixed value of $z_{\rm max}$. For protons, this parameter is largely irrelevant provided $z_{\rm max} \gtrsim 1$, but for the neutrino flux, it can be more significant. We have adopted the same value as the KM3NeT analysis to allow for direct comparison, but in practice, its effect is mostly degenerate with the source evolution parameter $m$. Importantly, our findings confirm previous results showing that a positive, rapidly evolving source population up to high redshift ($z \gg 1$) is needed to explain the observed neutrino flux.

Additionally, our model assumes protons as the only primary cosmic-ray species, neglecting the contribution of heavier nuclei. More comprehensive models, such as~\cite{PierreAuger:2022atd}, account for the LE and HE UHECR source populations including nuclei and their extragalactic disintegration, thereby altering the production of cosmogenic particles. In the energy range relevant to this study, cosmogenic neutrinos originate both from a sub-ankle population of primary protons with a steep spectrum (weakly constrained in terms of maximum energy) and from secondary protons produced by nuclei above the ankle. 
Consistently, our approach yields an \emph{effective} proton population that reproduces the observed local flux, requiring roughly twice the emissivity and a slightly harder spectrum than the sub-ankle component in Ref.~\cite{PierreAuger:2022atd}; these adjustments exert only a negligible influence on our final results.

%In conclusion, our results, based on a simple yet robust framework, clearly demonstrate that, given the observed energy, the cosmogenic origin of KM3-230213A can be attributed (as previously suggested) to either of two standard proton populations: a LE sub-ankle component with maximum energy around $5 \times 10^{18}$ eV, or a sub-dominant hard-spectrum HE component with GZK-scale maximum energy, both consistent with current UHECR composition data. In both cases, a cosmological evolution of $m \sim 3$ is required.

Crucially, our multimessenger analysis shows that the LE scenario, while compatible with the neutrino flux, would saturate or overshoot the diffuse gamma-ray background due to the expected cosmogenic gamma emission, rendering it less plausible as the origin of the observed event.
If, as our results suggest, only HE protons -- with hard spectra and high maximum energies -- can plausibly account for the observed event without violating gamma-ray constraints, then measuring the cosmic-ray composition above the ankle becomes critically important. Improved composition measurements, such as those anticipated from AugerPrime~\cite{Castellina:2019irv}, will be essential for testing the viability of the HE proton scenario and further constraining the origins of the highest-energy neutrinos in the Universe.

%\CE{Possiamo dire qualcosa sui campi magnetici?}

Additionally, this scenario highlights the importance of high energy neutrino observations ($E\gtrsim 10$~PeV), where running km3-scale detectors rapidly lose sensitivity. Next generation of neutrino observatories - including GRAND, RNO-G, POEMMA, IceCube-Gen2 - expected to improve the current sensitivity by at least an order of magnitude~\cite{2025JInst..20P4015A,GRAND:2018iaj,IceCube-Gen2:2020qha,Kotera:2021hbp,POEMMA:2020ykm}, will be instrumental to test the scenario presented in this paper.  By complementing the information provided by UHECR observatories, these future facilities will enable even more stringent constraints on the cosmogenic origin of the KM3-230213A event. 
%%%%%%%%%% END SECTION %%%%%%%%%%

%%%%%%%%%% BEGIN SECTION %%%%%%%%%%
\section{Methods}
\label{sec:methods}

\subsection*{Propagation of UHE Protons}

The calculation of the UHE proton spectra for each astrophysical scenario follows established analytic methods as in~\cite{Berezinsky:2002nc}. 
We assume that UHE protons are continuously injected throughout cosmic history, up to a maximum redshift $z_{\rm max} = 5$, with a power-law energy distribution and an exponential cutoff. 
The comoving injection rate per unit volume~\footnote{Throughout this section, comoving quantities are expressed in \textbf{boldface}, while proper quantities are indicated in normal font.} is expressed as:
\begin{equation}
\comoving{Q}_p(E,z) = \comoving{Q}_0 \left(\frac{E}{E_0}\right)^{-\gamma} \exp\left(-\frac{E}{E_{\rm max}}\right) f(z),
\end{equation}
where $E$ is the proton energy, $E_0 = 10^{17}$ eV is the normalization energy, $\gamma$ is the spectral index, and $E_{\rm max}$ is the maximum acceleration energy for protons at the source. The function $f(z) = (1 + z)^m$ describes the redshift evolution of the source population, with the evolution parameter $m$ explored over a representative range (see below).
The normalization constant $\comoving{Q}_0$ is fixed by the emissivity ${\comoving{L}}$ (the total energy injected per unit volume and time above $E_0$), as listed for each scenario in Table~\ref{tab:set_parameter}:
\[
\comoving{Q}_0 = \frac{\comoving{L}}{\int_{E_0}^{\infty} dE'\, E' \left(\frac{E'}{E_0}\right)^{-\gamma} \exp\left(-\frac{E'}{E_{\rm max}}\right)}.
\]

We consider several physically motivated choices for the source evolution parameter $m$. Strong source evolution ($m \gtrsim 4$) may correspond to luminous non-jetted active galactic nuclei or blazars~\cite{Ambrosone:2024zrf, Ajello:2015mfa}; intermediate values ($m \simeq 3$) reflect the cosmic star formation history~\cite{Ambrosone:2020evo}; while a flat distribution ($m \simeq 0$) is characteristic of intermediate-luminosity blazars~\cite{KM3NeT:2025vut}. 
This range captures the astrophysical uncertainty in UHECR source populations.

The equilibrium comoving density of UHE protons, $\comoving{n}_{p}(E, z)$, at redshift $z$ is computed by integrating the time-dependent transport equation, accounting for all relevant energy loss processes in the expanding Universe:
\begin{equation}
\comoving{n}_{p}(E, z) = \int_{z}^{z_{\rm max}} dz_g \left| \frac{dt}{dz_g} \right| \, \comoving{Q}_{p} \left( E_{g}(E, z, z_g), z_g \right) \frac{dE_g}{dE}(E, z, z_g)~.
\end{equation}

Here, $E_g(E, z, z_g)$ is the energy at emission (\emph{generation energy}) for a proton observed with energy $E$ at redshift $z$, taking into account energy losses due to interactions and cosmological redshift. The cosmological time-redshift relation is $\left| \frac{dt}{dz} \right| = 1/[H(z) (1 + z)]$, with $H(z)$ the Hubble parameter at redshift $z$. All calculations assume a standard flat $\Lambda$CDM cosmology~\cite{Planck:2018vyg}.

The intensity of UHE protons at Earth (i.e., $z = 0$) is then given by~\cite{Berezinsky:2002nc}:
\begin{equation}
I_p(E) = \frac{c}{4\pi} \int_0^{z_{\rm max}} dz_g \left| \frac{dt}{dz_g} \right| \, \comoving{Q}_{p}(E_{g}(E, 0, z_g), z_g) \frac{dE_g}{dE}(E, 0, z_g).
\end{equation}

\subsection*{Calculation of Cosmogenic Neutrino and Gamma-ray Fluxes}

The rate of production of secondary particles (such as neutrinos and gamma rays) from protons interacting with photon backgrounds is determined by the equilibrium proton density. For photopion production, the injection rate of secondary species $i$ (where $i = \nu,\, \gamma,\, e^{\pm}$) per unit comoving volume is:
\begin{equation}
\comoving{Q}_i^\pi (E_i, z) = \int_{E_i} \frac{dE_p}{E_p} \, \comoving{n}_p(E_p, z) \int_{\epsilon_{\rm th}(E_p)} d\epsilon \, n_b(\epsilon, z) \, \Phi(E_i, E_p, \epsilon).
\end{equation}
Here, $n_b(\epsilon, z)$ is the total spectral density of background photons, including both the cosmic microwave background (CMB) and the extragalactic background light (EBL).
%
%$$
%n_b(\epsilon, z) = n_{\rm CMB}(\epsilon, z) + n_{\rm EBL}(\epsilon, z).
%$$
%
The function $\Phi$ is the differential cross-section for photohadronic interactions, for which we use the parametrization of~\cite{Kelner:2008ke}, based on the \texttt{SOPHIA} Monte Carlo code~\cite{Mucke:1999yb}. The photopion production threshold energy, $\epsilon_{\rm th}(E_p)$, is determined by kinematics.

For the CMB, the photon density evolves as
\[
n_{\rm CMB}(\epsilon, z) = (1+z)^2 n_{\rm CMB}^{0}\left(\frac{\epsilon}{1+z}\right),
\]
where $n_{\rm CMB}^{0}$ is the present-day CMB blackbody spectrum. For the EBL, we adopt the recent model provided by~\cite{Saldana-Lopez:2020qzx}.

The cosmogenic neutrino intensity at Earth is calculated similarly to protons but accounting only for adiabatic (cosmological) energy losses, as neutrinos propagate essentially unabsorbed:
\begin{equation}
I_\nu (E) = \frac{1}{3}\frac{c}{4\pi} \int_0^{z_{\rm max}} dz_g \left| \frac{dt}{dz_g} \right| \comoving{Q}_\nu(E(1+z_g), z_g) (1 + z_g),
\end{equation}
where $\comoving{Q}_\nu(E, z)$ is the sum over all neutrino flavors produced in photopion interactions, and it is computed at the neutrino energy  $E(1+z)$ to take redshift into account. 
The factor $1/3$ accounts for neutrino flavor oscillations during propagation from source to Earth.
For gamma rays, we assume that electromagnetic cascades initiated by high-energy photons or electrons/positrons rapidly lose energy and reprocess it into lower-energy photons, effectively ``recycling'' the initial injected energy within the cascade. Following~\cite{Berezinsky:2016feh}, the injection rate of cascade gamma rays is modeled as:
\begin{equation}
\comoving{Q}_\gamma(E, z) = \frac{\comoving{\omega}_{\rm casc}(z)}{\epsilon_X^2[2 + \ln(\epsilon_C/\epsilon_X)]}
    \begin{cases}
        \left(E/\epsilon_X\right)^{-3/2} & \text{if } E \lesssim \epsilon_X~, \\
        \left(E/\epsilon_X\right)^{-2}    & \text{if } E \gtrsim \epsilon_X~,
    \end{cases}
\end{equation}
where $\epsilon_X$ and $\epsilon_C$ are the characteristic (``break'' and ``cutoff'') energies for cascade development, and $\comoving{\omega}_{\rm casc}(z)$ is the total cascade energy injection rate at redshift $z$. For the calculation, we approximate the CMB and EBL as monochromatic backgrounds with energies corresponding to their respective peaks ($\epsilon_{\rm CMB}(z) = k_B~T(z)$, $\epsilon_{\rm EBL}(z) = 0.68$~eV).

The total energy injected into electromagnetic cascades is computed separately for contributions from photopion:
\begin{equation}
\comoving{\omega}_{\rm casc}^\pi(z) = \int dE \, E \left[\comoving{Q}_{e^-}^\pi(E, z) + \comoving{Q}_{e^+}^\pi(E, z) + \comoving{Q}_{\gamma}^\pi(E, z)\right],
\end{equation}
and pair production processes:
\begin{equation}
\comoving{\omega}_{\rm casc}^{ee}(z) = \int dE_p \, E_p^2 \, \comoving{n}_p(E_p, z) \, \beta(E_p, z),
\end{equation}
where $\beta(E_p, z)$ is the energy loss rate of protons due to pair production on the CMB and EBL~\cite{Chodorowski1992ApJ}.

The observable diffuse gamma-ray intensity at Earth is then
\begin{equation}
I_\gamma(E) = \frac{c}{4\pi} \int_0^{z_{\rm max}} dz_g \left| \frac{dt}{dz_g} \right| \comoving{Q}_\gamma \left( E(1+z_g), z_g \right) (1+z_g).
\end{equation}

\subsection*{Parameter sets of tested models}

All tested scenarios for the low-energy (LE) and high-energy (HE) source populations, as well as the resulting total cascade energy densities, are summarized in Table~\ref{tab:set_parameter}. Model calculations employ updated cosmological parameters from Planck 2018~\cite{Planck:2018vyg}.

\begin{table}[h!]
\caption{Parameter sets adopted for this analysis. From left to right, the columns report: the source redshift evolution parameter ($m$), spectral index ($\gamma$), maximum proton energy ($E_{\rm max}$), total proton luminosity ($\comoving{\mathcal{L}}$), and the integrated contributions to the electromagnetic cascade energy density from proton pair production ($\Omega_{\rm tot}^{ee}$) and photopion interactions ($\Omega_{\rm tot}^\pi$). The upper and lower sections correspond to the low-energy (LE) and high-energy (HE) source populations, respectively.}\label{tab:set_parameter}
\centering
\begin{tabular}{|c|c|c|c|c|c|c|}
\hline
& m & $\gamma$ & $E_{\rm max}\, [10^{18}\, \rm eV]$ & $\comoving{L} \, [10^{45} \frac{\rm erg}{\rm Mpc^3 yr}]$ & $\Omega_{\rm tot}^{ee} \, [\frac{\rm eV}{\rm cm^3}]$ & $\Omega_{\rm tot}^{\pi} \, [\frac{\rm eV}{\rm cm^3}]$ \\
\hline
\multirow{3}{*}{\rotatebox{90}{\textbf{LE}}} & 
{\color{Red}0} & {\color{Red}2.5} &  {\color{Red}6.0}  & {\color{Red}2.68} & {\color{Red}$2.90 \times 10^{-9}$} & {\color{Red}$1.31 \times 10^{-10}$} \\ 
\cline{2-7} & 
{\color{Green}3} & {\color{Green}2.01} & {\color{Green}5.0} & {\color{Green}17.3} & {\color{Green}$3.09 \times 10^{-8}$} & {\color{Green}$1.17 \times 10^{-9}$} \\ 
\cline{2-7} & 
{\color{blue}5} & {\color{blue}1.4} & {\color{blue}4.0} & {\color{blue}0.24} & {\color{blue}$2.22 \times 10^{-7}$} & {\color{blue}$6.85 \times 10^{-9}$} \\ 
\cline{2-7}
\hline
\multirow{3}{*}{\rotatebox{90}{\textbf{HE}}} & 
{\color{Red}0} & {\color{Red}1.3} & {\color{Red}$10^2$} & {\color{Red}$1.59\times 10^{-2}$} & {\color{Red}$1.47 \times 10^{-10}$} & {\color{Red}$6.02 \times 10^{-11}$} \\ 
\cline{2-7} & 
{\color{Green}3} & {\color{Green}1.0} & {\color{Green}$10^2$} & {\color{Green}$9.86\times 10^{-3}$} & {\color{Green}$8.46 \times 10^{-10}$} & {\color{Green}$6.30 \times 10^{-10}$} \\ 
\cline{2-7} & 
{\color{blue}5} & {\color{blue}0.7} & {\color{blue}$10^2$} & {\color{blue}$6.86\times 10^{-3}$} & {\color{blue}$5.52 \times 10^{-9}$} & {\color{blue}$6.37 \times 10^{-9}$} \\ 
\hline
\end{tabular}
\end{table}

\bibliographystyle{apsrev4-2}
\bibliography{biblio}

\newpage

% !TEX root = ./main.tex
\section*{Data Availability}

All data analyzed in this study are publicly available from the sources cited in the manuscript. Cosmic-ray data were obtained from the Cosmic-Ray DataBase (CRDB) v4.1~\cite{Maurin:2023alp}~\footnote{\url{https://lpsc.in2p3.fr/crdb}}, and from the KASCADE Cosmic Ray Data Centre (KCDC)~\footnote{\url{https://kcdc.iap.kit.edu/}}. Additional datasets referenced in this work are available from the corresponding publications and data repositories as cited.

%as the all-particle Auger spectrum \cite{PierreAuger:2021hun}, while the Auger-proton spectrum is obtained multiplying the all-particle spectrum by the proton fraction obtained in \cite{PierreAuger:2023xfc}. The KASCADE-Grande and IceCube proton spectra are respectively available in~\cite{Apel:2013uni,IceCube:2020yct}. The IceCube neutrino diffuse fluxes are available in~\cite{IceCube:2020acn,Abbasi:2021qfz,IceCube:2021rpz}, while the KM3-230213A equivalent flux can be found in Ref.~\cite{KM3NeT:2025ccp}. Finally, the IGRB flux is reported in \cite{Fermi-LAT:2014ryh}.}

\section*{Code Availability}

The code used in order to produce the results presented in this manuscript can be accessed upon motivated request to the corresponding author.

\section*{Acknowledgements}

AA acknowledges the support of the project ``NUSES - A pathfinder for studying astrophysical neutrinos and electromagnetic signals of seismic origin from space'' (Cod.~id.~Ugov: NUSES; CUP: D12F19000040003). The work of RA and CE has been partially funded by the European Union – Next Generation EU, through PRIN-MUR 2022TJW4EJ and by the European Union – NextGenerationEU under the MUR National Innovation Ecosystem grant ECS00000041 – VITALITY/ASTRA – CUP D13C21000430001. 

\section*{Author Contributions}

AC led the calculations and the analysis and he is the primary author of the article. All the authors equally contributed to the idea developments, the interpretation of the results, the writing and to the editing of the manuscript.

\section*{Competing Interest}

The authors declare no competing interests.

\end{document}